# Ion beam induced enhanced diffusion from gold thin films in silicon

J. Ghatak[1], B. Sundaravel[2], K. G. M. Nair[2], and P. V. Satyam[1*]

[1]Institute of Physics, Bhubaneswar - 751005, India
[2]Materials Science Division, Indira Gandhi Center for Atomic Research, Kalpakkam - 603102, India

## Abstract

Enhanced diffusion of gold atoms into silicon substrate has been studied in Au thin films of various thicknesses (2.0, 5.3, 10.9 and 27.5 nm) deposited on Si(111) and followed by irradiation with 1.5 MeV $Au^{2+}$ at a flux of $6.3 \times 10^{12}$ ions $cm^{-2}$ $s^{-1}$ and fluence up to $1 \times 10^{15}$ ions $cm^{-2}$. The high resolution transmission electron microscopy measurements showed the presence of gold silicide formation for the above-mentioned systems at fluence $\leq 1 \times 10^{14}$ ions $cm^{-2}$. The maximum depth to which the gold atoms have been diffused at a fluence of $1 \times 10^{14}$ ions $cm^{-2}$ for the cases of 2.0, 5.3, 10.9 and 27.5 nm thick films has been found to be 60, 95, 160 and 13 nm respectively. Interestingly, at higher fluence of $1 \times 10^{15}$ ions $cm^{-2}$ in case of 27.5 nm thick film, gold atoms from the film transported to a maximum depth of 265 nm in the substrate. The substrate silicon is found to be amorphous at the above fluence values where unusually large mass transport occurred. Enhanced diffusion has been explained on the basis of ion beam induced, flux dependent amorphous nature of the substrate, and transient beam induced temperature effects. This work confirms the absence of confinement effects that arise from spatially confined structures and existence of thermal and chemical reactions during ion irradiation.

PACS: 61.80.Jh, 61.82.-**d**, 68.35.-p, 68.37.Lp, 79.20.Rf

Keywords: Gold nanostructures, MeV ion irradiation, diffusion, segregation, amorphous silicon

# 1. Introduction

The study of diffusion of noble metals in crystalline silicon has been actively pursued due to various technological applications and associated fundamental issues. As the electronic device size shrinking to nanometer regime, a better understanding of the effect of material interfaces at nanoscale regime on atomic processes is necessary. Gold and Silicon are two of the commonly used elements in semiconductor production. One of the applications of this combination is related to the Au–Si eutectic bonding, which is used in integrated circuits packaging and micro-electro-mechanical system (MEMS) applications, such as for sealing, die attachment, and electrical interconnects [1]. Gold is widely used in electronic devices to control the base minority-carrier life time due to its ability to act as a recombination centre when dissolved in silicon [2]. In the fabrication of such devices, gold is introduced into silicon wafers by high temperature diffusion process following vacuum evaporation of metallic gold on the surfaces. It is also well known that gold thin films act as contact layers in many systems for electronic measurements. Hence, the diffusion of gold atoms in crystalline silicon play important role. Two mechanisms, namely, "kick-out" mechanism [3] and "Frank-Turnbull" mechanism [4] have been used to explain the gold atoms diffusion in crystalline silicon. Similar understanding of diffusion of gold in amorphous silicon ($a$–Si) is not well understood [5,6].

The study of the diffusion of gold atoms in $a$–Si at high temperatures would be difficult as the recrystallization of silicon plays a role and hence it is not possible to study diffusion phenomena in amorphous silicon under high temperature thermal equilibrium conditions. Past investigations of Au-implanted silicon has suggested about the gold segregation as a result of its being expelled from the recrystallized amorphous layer during thermal and ion beam annealing [7, 8]. As the crystalline silicon matrix is being amorphized the gold solid solubility and diffusivity changes. It has been shown that the solid-solubility in crystalline silicon ($\approx 10^{11}$ Au atoms cm$^{-3}$) is six orders of magnitude *lower* than that in amorphous silicon [9,10]. Similarly, the diffusivity of Au in crystalline silicon is lower by many orders of magnitude compared the diffusion in amorphous silicon [10,11]. In the study of precipitation of implanted atoms, both the segregation and diffusion of gold atoms has been found to play a role: the segregation into a densely defected region where it exceeds the local solubility resulting in precipitation and the diffusion along the dislocations network until a node where again it exceeds the local threshold for precipitation [12]. Alford et al also showed that the enhanced diffusion depends upon the magnitude of dynamic recrystalization occurring during implantation [12]. Recently, Hedler et al., demonstrate that the amorphous silicon deforms plastically in the same way as conventional glasses during irradiation with high energy heavy ions and provided an experimental evidence for the existence of the low-density liquid [13].

In our earlier studies, interesting differences for *ion–uniform film* and *ions–nanostructure* interaction phenomena, particularly relating to ion beam mixing and surface morphological effects have been reported [14 – 17]. It was found out that under the low – flux ($\approx 1.3 \times 10^{11}$ ions cm$^{-2}$ s$^{-1}$) irradiation conditions, ion beam induced mixing is found to be absent for continuous Au film on silicon compared to isolated nanostructure systems [15]. For this case, mixing at low flux was observed only in case of grazing incidence irradiation geometry (impact angle of 60º) but *no enhanced* diffusion was observed [15]. One of the main issues that were intriguing in nature was the role of spatial confinement of nanostructures in comparison with thicker and continuous films which was addressed in ref [15]. In the present work, the thickness dependent studies (i.e., mass transport under high flux condition from various sizes of gold nanostructures, from isolated morphology to uniform and continuous morphology) confirmed that the mass transport is not as a consequence of the direct cascade induced flow and the spatial confinement of the nanostructure present on the surface and but the mechanism would be proposed based on the amorphous nature of the substrate, ion flux and fluence to explain the enhanced diffusion occurring for gold atoms in *amorphous*-silicon system.

## 2. Experimental Methods

Au films of thickness ≈2.0, 5.3, 10.9 and 27.5 nm thick Au films were deposited at room temperature by thermal evaporation using resistive heating method in high vacuum conditions($\approx 4 \times 10^{-6}$ mbar) on a ≈2 nm thick native oxide covered Si(111) substrates. The thickness has been determined using Rutherford back-scattering spectrometry (RBS) measurements. The substrates were cleaned with de-ionized water followed by rinsing in methanol and acetone prior to deposition. Deposition rate for 2.0 and 5.3 nm samples was kept at ≈ 0.01 nm/s and for 10.9 and 27.5 nm kept at ≈ 0.1 nm/s. All the irradiations were carried out with 1.5 MeV Au$^{2+}$ ions at room temperature with ion flux $6.3 \times 10^{12}$ ions cm$^{-2}$ s$^{-1}$ (corresponding current density is 2.0 μA cm$^{-2}$) using the 1.7 MV Tendetron accelerator at Indira Gandhi Center for Atomic Research, Kalpakkam. The fluences on the samples were varied from $1 \times 10^{14}$ to $1 \times 10^{15}$ ions cm$^{-2}$. The substrates were oriented 5$^0$ off normal to the incident beam to suppress the channeling effect and mounted using silver dug on bulk copper holder. The RBS measurements were carried out with 1.35 MeV He$^+$ ions using 3 MV pelletron at Institute of Physics, Bhubaneswar. Spectra are collected with surface barrier detector kept at an angle of 160$^0$ placed ≈ 10 cm away from the samples. Cross-sectional transmission electron microscopy (XTEM) measurements were performed at Institute of Physics, Bhubaneswar using JEOL JEM-2010 operating at 200 kV on the irradiated samples to study the surface and interface modifications. XTEM samples were prepared by mechanical thinning followed by 3.5 keV Ar ion milling. It is to be noted that projected ranges of 1.5 MeV Au ions in Au and Si are 96 and 320 nm [18] and hence the implanted species would penetrate deep into Si substrate, much more than the average height of the Au layer for all the cases.

## 3. Experimental results

Recently, a direct observation of dramatic mass transport due to 1.5 MeV $Au^{2+}$ ion impact on isolated Au nanostructures of an average size $\approx$ 7.6 nm and a height $\approx$ 6.9 nm that are deposited on Si (111) substrate under high flux ($3.2\times10^{10}$ to $6.3\times10^{12}$ ions $cm^{-2}$ $s^{-1}$) conditions has been reported [17]. The nanostructures were formed as a result of non-wetting nature of thin gold film of thickness $\approx$ 2.0 nm on the native oxide covered surface. The cross-sectional TEM measurements showed maximum mass transport extending up to a distance of about 60 nm into the substrate, much beyond the size of the nanostructures. The unusual mass transport has been found to be associated with the formation of gold silicide nanoalloys at sub-surfaces. The confirmation of the presence of Au atoms transport from the nanostructures following the irradiation has been obtained using the Scanning TEM – high angle annular dark field (HAADF) imaging method [17]. In the present work, results on other thickness films (resulting in isolated gold structures to uniform and continuous films). The results show the importance of the substrate structure (crystalline or amorphous) and the availability of gold atoms that are recoiled or ballistically mixed at the interface in the initial stages of the irradiation.

Figures in 1(a) - (b) are the BF XTEM images of pristine 5.3 and 10.9 nm Au on Si (111) substrates. For the pristine film of thickness 5.3 nm, XTEM images show large isolated nanostructures as shown in Figure 1 (a), while for the film thickness of 10.9 nm, the uniformity and the coverage area of gold nanostructures increases. The above thickness values of gold films determined using RBS measurements and followed by SIMNRA simulation [19]. During the simulations, bulk density of Au has been used for obtaining the thickness values. The irradiation effects at a fluence $1\times10^{14}$ ions $cm^{-2}$ with flux $6.3\times10^{12}$ ions $cm^{-2}$ $s^{-1}$ are shown in Figures 1(c) and (d) for 5.3 and 10.9 nm gold on Si(111) samples, respectively. In these two cases, enhanced diffusion has been observed from the top gold layer into the Si substrate. For the case of 5.3 nm Au/Si, the maximum depths for Au atoms diffusion

in to Si substrates has been found to be ≈ 95 nm. Irradiation effects in 10.9 nm thick Au on Si system shows the more mass transport into Si than both the 2.0 nm and 5.3 nm thick gold film cases. For 10.9 nm thick film case, the maximum transported depth was found to be as large as ≈160 nm (Figure 1(d)). The contrast that is seen in the bright field images (Fig. 1(c) and (d)) has been attributed to the contribution from the diffused gold atoms (i.e. presence of gold atoms at larger depth in the interfacial region). This contrast is same as that was observed for 2.0 nm Au/Si systems, for which the confirmation has been done using STEM – HAADF imaging methods [15]. These observations reveal that the amount of mass transport (and maximum depth) increases with the increase of top Au layer thickness ruling out the confinement effects that are due to size of the nanostructures. It is also important to note the crystalline nature of the silicon substrate play an important role as the diffusion of gold atoms in amorphous silicon is many orders magnitude higher than in crystalline system [10,11]. Figure 1(e) corresponds to the selected area electron diffraction (SAED) pattern taken from the implanted region (circular region of Figure 1(d)) of 5.3 nm Au on Si system and this confirms the amorphization of the Si substrate at this fluence. This is also the same case with the irradiated system of 10.9 nm thick film case. Hence, the diffusion that has been observed is due to the amorphous nature of silicon substrate. In the following we show that, even if the gold atoms are present at the interface, and the substrate is not completely amorphized, then the diffusion lengths would be limited. Figure 1(f) is taken from the rectangular region (near surface) of Figure 1(d), for irradiated 10.9 nm Au/Si system. The selected region show crystalline in nature with the lattice spacing corresponding to 0.304 ± 0.005 nm. This indicates the formation of hexagonal $Au_5Si_2$ phase [15, 17, 20]. The gold silicide phase formation indicates that the mass transport is also associated with the thermal reaction. More details about the silicide formation have been reported elsewhere [17].

The role of the crystalline nature of the substrate for the enhanced diffusion of gold atoms is evident from the mass transport studies from a relatively thicker gold film (of thickness 27.5 nm thick on silicon). Figures 2(a) depicts the bright field (BF) XTEM micrographs for pristine Au film thickness of 27.5 nm. RBS measurements using 1.35 MeV $He^+$ for as-deposited 27.5 nm Au/ silicon substrate is shown in Fig. 2(e). The thickness has been determined using the RBS measurements and is in agreement with the cross-sectional TEM as shown in Figure 2(a). The TEM measurements also show that the morphology is a continuous one unlike the thin film case (2 nm thick). Figures 2(b) and (c) are the XTEM micrographs of the irradiated 27.5 nm Au/Si systems with a beam flux $6.3 \times 10^{12}$ ions $cm^{-2}$ $s^{-1}$ at fluence $1 \times 10^{14}$ and $1 \times 10^{15}$ ions $cm^{-2}$ respectively. Interestingly, for the case of 27.5 nm Au/Si system, we have observed a mass transport up to a maximum depth of only ≈13 nm (see figure 2(b)) at a fluence $1\times10^{14}$ ions cm$^{-2}$. It is to be noted that, at this fluence ($1\times10^{14}$ ions cm$^{-2}$) and flux ($6.3\times10^{12}$ ions cm$^{-2}$ s$^{-1}$), a large mass transport was observed in thinner films (2.0, 5.3 and 10.9 nm thick Au/Si systems). Previous low flux measurements ($1.3\times10^{11}$ ions cm$^{-2}$ s$^{-1}$) at the same fluence show *no mass transport and mixing* at the interface [15]. The mass transport has been increased dramatically up to 265 nm for this case at fluence $1\times10^{15}$ ions cm$^{-2}$ as shown in figure 2(c). The inset of figure 2(c) is the SAED from the implanted region of the substrate and indicates that the substrate up to projected range is completely amorphous. The lattice spacing of high resolution TEM image taken from the rectangular region of figure 2(c) again indicates the formation of Au-Si alloy as mentioned earlier for thin film cases. From the high-resolution XTEM measurements for all the above-mentioned samples, we conclude that high flux MeV ion irradiation leads to alloy formation for various thickness of Au films on Si which was absent during the irradiation at low flux ($1.3\times10^{11}$ ions cm$^{-2}$ s$^{-1}$) conditions [15].

As the mass transport in case of 27.5 nm sample is less in depth at a fluence $1\times10^{14}$ ions cm$^{-2}$, to confirm Au transportation into Si we performed the XTEM and RBS measurements on aqua regia ($HNO_3$ : HCL = 1:3) treated irradiated sample. After irradiation, we have etched the top Au layer with aqua regia which is known to etch pure Au and then performed XTEM and RBS on aqua regia treated samples. Figure 2(d) is the BF XTEM of the after aqua regia treated sample which was irradiated with fluence value of $1\times10^{14}$ ions cm$^{-2}$ and confirms the absence of Au in top surface with the presence of Au-Si alloy embedded in Si (aqua regia does not etch gold silicide alloy). Figure 2(e) is the RBS spectra for pristine, as-irradiated and aqua regia etched samples. For the sake of visualization, only Au signal for aqua regia treated sample has been magnified (multiplied by a factor of 10). From the spectra it is evident that there is presence of Au even after Aqua regia treatment. The right shift of Si edge and left shift of Au edge for the aqua regia treated sample reflects the absence of Au layer on Si surface but the presence if Au as embedded in to the Si, in agreement with XTEM measurements. As mentioned earlier, the alloy formation has been confirmed using lattice imaging from the TEM measurements.

We now discuss the role of recoil events to understand the observed experimental data. Using SRIM simulations, the energy required for the redistributed Au atoms to reach a depth of 160 nm into Silicon would be about ≈ 600 keV [16]. The nuclear energy loss and the electronic energy loss for 1.5 MeV Au ions in Au are 9.5 keV/nm and 2.5 keV/nm, respectively. From the total energy loss for 1.5 MeV Au ions in 10.9 nm Au/Si systems thick gold target is only about 131 keV, which is less than the energy required to go 160 nm inside the Si. Assuming that this is the maximum energy available for forward Au atom (from the Au layer), it is *not* possible to explain the depth that is achieved by forward moving Au atoms. Again SRIM simulation with 1.5 MeV Au ion bombardments on 10.9 nm thick Au film on Si substrate showed that the maximum recoil depth for the Au atoms into Si arising from the top film found to be ≈ 3.5 nm from the interface and is found to be independent of top Au film thickness. It should be noted that SRIM simulations are done at 0 K (no thermal effects are taken into account) whereas our experiment has been carried out in room temperature. There by we cannot explain the huge mass transport, just simply by recoil event and thermal effect along with beam flux induced effects have to be taken into account. Similarly the unusual mass transport observed for 27.5 nm thick film at a flux value of $1\times10^{15}$ ions cm$^{-2}$ (265 nm) can't be explained by the recoil events alone. But, mass transported depth of 13 nm for high flux and low fluence ($1\times10^{14}$ ions cm$^{-2}$) can be considered as recoil or ballastically mixed events only.

Recoil of target atoms is also a pronounced effect due to ion bombardment and it is an instantaneous process as it is a cascade mediated phenomena. From the present and earlier experimental results [17], it reveals that recoil of Au atoms from the top Au layer is highly influenced by the ion beam flux and the nature of substrate (in terms of defect concentration) and at constant fluence, there is a critical flux exists below which recoil does not noticeable. To get more recoil one has to increase the deposited energy by changing beam impact angle [15, 18], by increasing mass and energy of ion [16] and by increasing fluence. As we have already explained, using SRIM simulations [18] the energy required for the redistributed Au atoms to reach a depth of present experimental values come out to be less than the available energy from the energy loss due to the 1.5 MeV Au ion beam [17]. Again TRIM simulation with 1.5 MeV Au ion bombardments on 10.9 nm thick Au film on Si substrate showed that the maximum recoil depth for the Au atoms into Si arising from the top film found to be ≈ 3.5 nm from the interface and is found to be independent of top Au film thickness. There by we can not explain the mass transport just simply by recoil of Au atoms and this indicates to the diffusion phenomenon and one has to take the high flux effect into account.

It is known that in semiconductor, defect concentration increases with ion flux which causes the amorphization faster (at lower fluence with higher flux) [21]. That means at a given temperature (around room temperature), with higher flux, lower fluence is needed to achieve amorphization whereas with lower flux, the defect concentration has to be compensated with higher fluence [22]. From the previous flux dependent study, the amorphization has been observed at fluence $6\times10^{13}$ ions cm$^{-2}$ with ion flux $6.3\times10^{12}$ ions cm$^{-2}$ s$^{-1}$ [17] which is less compared to the earlier observation [23] and hence we may infer that the higher flux leads to the faster amorphization in present case also. This is so as the displacements production (vacancy-interstitial pair) rate is proportional to the incident ion flux [24]. In the present experimental condition beam heating is also a crucial factor for the mass transport as it mediates through diffusion. We have estimated the wafer temperature during irradiation [17] using the experimentally verified prescription given by Nakata [25]. For the flux $6.3 \times 10^{12}$ ions cm$^{-2}$ s$^{-1}$, the temperature rises due to fluence $1 \times 10^{14}$ (irradiation time $\approx$15 sec) is 1150 K and at a fluence $1 \times 10^{15}$ ions cm$^{-2}$ ($\approx$150 sec) is 1150 and 1155 K respectively. It also turns out to be the higher temperature during irradiation for higher flux at a constant fluence.

One can divide the diffusion in to two regimes: temperature dependent (thermal diffusion) and temperature independent (radiation enhanced diffusion). In the temperature dependent regime, thermal diffusivity is proportional to the substrate temperature. There is a critical temperature ($T_c$) exists, above which the radiation enhanced diffusion (RED) will be significant [26]. It is to be noted that $T_c$ = 422 K for Au-Si case [27] which is less than the wafer temperature during irradiation due to beam heating and hence the phenomenon is strongly influenced by RED of recoiled Au atoms. Now the temperature of the wafer is high enough even when the substrate is partially amorphous (low flux irradiation or irradiation in higher thickness film) but the diffusivity of Au in amorphous Si (*a*-Si) is many orders more than that of crystalline Si (*c*-Si) [10,11]. This is due to the excess displacements created due to higher beam flux. Hence during irradiation, as soon as the amorphization takes place in Si substrate, the diffusivity of recoiled Au atoms increases drastically. A rough transported depth may be estimated from a diffusion length of marker layer model due to ballistic cascade mixing [28]. Considering the values of diffusivity $D_{cas} \approx 700$ nm$^2$ sec$^{-1}$ at T $\approx$1000K [11], average deposited energy $F_D(x)$ = 1761 eV nm$^{-1}$ [18], the transported depth (diffusion length) comes out to be $\approx$ 250 nm which is fairly matches with our observed transported depth. The difference in transported depth for different film thickness will be discussed in the following section.

From the results for various thicknesses of Au films, it is clear that the distributed depth of Au atoms is not similar for all the systems. For example, in case of 2 nm Au/Si, maximum depth of distributed Au is about 60 nm [17], whereas only 13 nm for 27.5 nm Au/Si. The same thing for 10.9 nm Au/Si system is as deep as 160 nm. The reasons could the following: In figure 3 we have plotted the SRIM simulation [18] of Au recoil distribution of 2, 5.3, 10.9 and 27.5 nm system due to 1.5 MeV Au at normal incidence. In the inset of figure 3, is only the zoomed recoil distribution of Au in to Si for all above-mentioned Au film thickness. It is clear from the figure that the amount of Au into Si (number of recoiled Au atoms per unit length of Si) is more in higher film thickness. So, if the available number of recoiled Au atoms to be diffused, are more, then the area of redistributed region also will be more and hence the transported depth also will be more. But the amorphization plays role in the diffusion process. For the same energetic ion beam, there will be different onset fluence for complete amorphization of Si substrates with Au films of different thicknesses. This is due to the fact that, in Si with thicker film, most of the energy would be transferred to over layer that the substrate and hence needs higher fluence for complete amorphization of the substrate. Similar conclusion is available from the experiment of Ehrhard et. al, with a thick Au layer deposited on *c*-Si, *a*-Si and partially *a*-Si [6]. It was clearly shown that Au diffusion from top Au layer is more in a-Si than in c-Si and any intermediate crystalline layer (that is partially a-Si) will suppress the effect. That is why we observed that the depth of mass transport is more at fluence $1\times10^{15}$ ions cm$^{-2}$ than that of $1\times10^{14}$ ions cm$^{-2}$ in case of 27.5 nm Au/Si systems. Hence the mechanism to understand the dramatic mass transport could be the following: (i) collision cascades drive gold atoms from the film into the substrate through ballistic mixing and recoiled implantation processes. (ii) This is followed by amorphization in the substrate that occurs at lower fluence at high flux condition. (iii) The effective wafer temperature due to incident beam power, drives the mass transport (enhanced diffusion occurring for gold atoms in *amorphous*-silicon system).

## 5. Conclusions

We experimentally demonstrated the unusual mass transport for Au nano-islands/Si systems due to MeV Au ion bombardment as a function of film thickness. The transported depth is more for higher thickness of Au film as we have observed maximum transported depth is about 265 nm for 27.5 nm Au/Si. Recoiled Au atoms from the top Au layer found to be more for higher thickness layer which is consistent with the SRIM simulation. Mass transport in all the samples starting with nanoislands to continuous thick layer suggests that the spike confinement is insignificant in this issue. Mass transport may be explained in the light of radiation enhanced diffusion of recoiled Au atoms into amorphous Si. High ion beam flux cause the faster amorphization and the wafer temperature rise which helps in faster diffusion of the recoiled Au atoms. Ion beam mixing can be achieved in thick continuous film also in case of Au-Si systems which was absent in earlier observation.

**Acknowledgements**

We acknowledge the cooperation of K. G. M. Nair, S. Amirthapandian and S. Santhanaraman during high current ion irradiation experiments. We would also like to thank all the staffs of both the beam hall at IOP and IGCAR.

**Figure Captions:**

**Figure 1:** Cross-sectional bright field TEM micrographs for pristine Au/Si systems: (a) 5.3 nm and (b) 10.9 nm Au/Si system. In all the samples ≈ 2.0 nm native oxide is present. (c) and (d) correspond to the irradiated 5.3 nm and 10.9 nm Au/Si system respectively. In all samples, irradiations were carried out carried with 1.5 MeV $Au^{2+}$ ions at a fluence of $1\times10^{14}$ ions $cm^{-2}$ with a flux $6.3\times10^{12}$ ions $cm^{-2}$ $s^{-1}$. (e) corresponds to the SAED pattern shows the amorphous nature of the implanted region of the substrate. (f) Corresponds to the high resolution cross sectional bright field TEM image of rectangular region of fig (d).

**Figure 2:** (Color online) (a) corresponds to the XTEM micrograph of pristine 27.5 nm Au/Si. (b) and (c) are the XTEM micrograph of 27.5 nm Au/Si after irradiated with 1.5 MeV $Au^{2+}$ ions at a fluence of $1\times10^{14}$ and $1\times10^{15}$ ions $cm^{-2}$ respectively. Beam flux was mentioned constant during irradiation with $6.3\times10^{12}$ ions $cm^{-2}$ $s^{-1}$. Inset of figure (c) depicts the SAED pattern after irradiation from the implanted region of the Si substrate. The high resolution XTEM image was taken from the rectangular region of fig (f) which is shown by arrow mark. (d) is the XTEM micrograph of aqua regia etched irradiated sample while the fluence was $1\times10^{14}$ ions $cm^{-2}$. (e) corresponds to the RBS spectra of irradiated 27.5 nm Au/Si system at before and after aqua regia treatment.

**Figure 3:** (Color online) SRIM profile for Au recoil distribution due to 1.5 MeV Au ions. The Au thickness have been used in the simulation are 2, 5, 10 and 27.5 nm. Inset corresponds to the zoomed version of the recoil distribution in to Si for all above mentioned thickness of Au films.

**Figure 1: J. Ghatak et al**

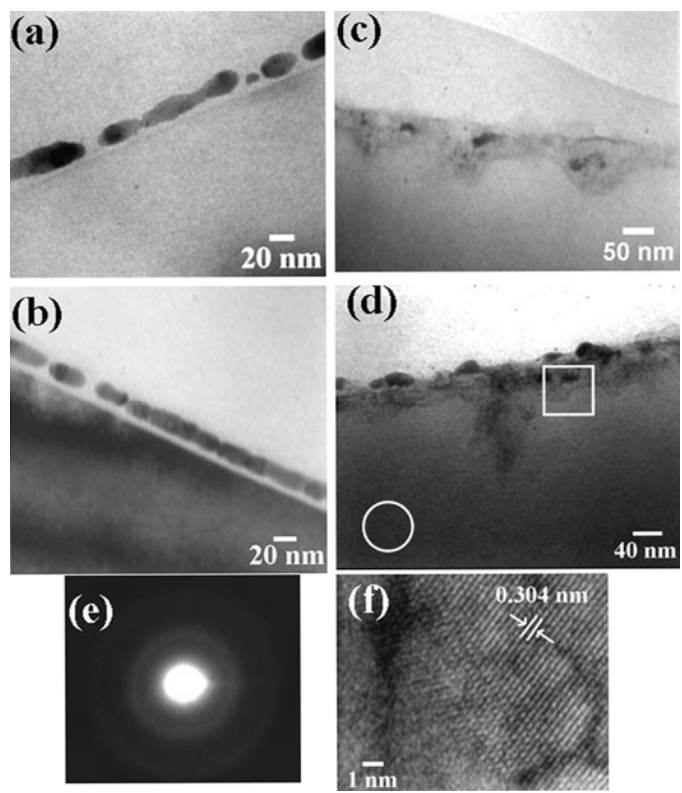

**Figure 2: J. Ghatak et al**

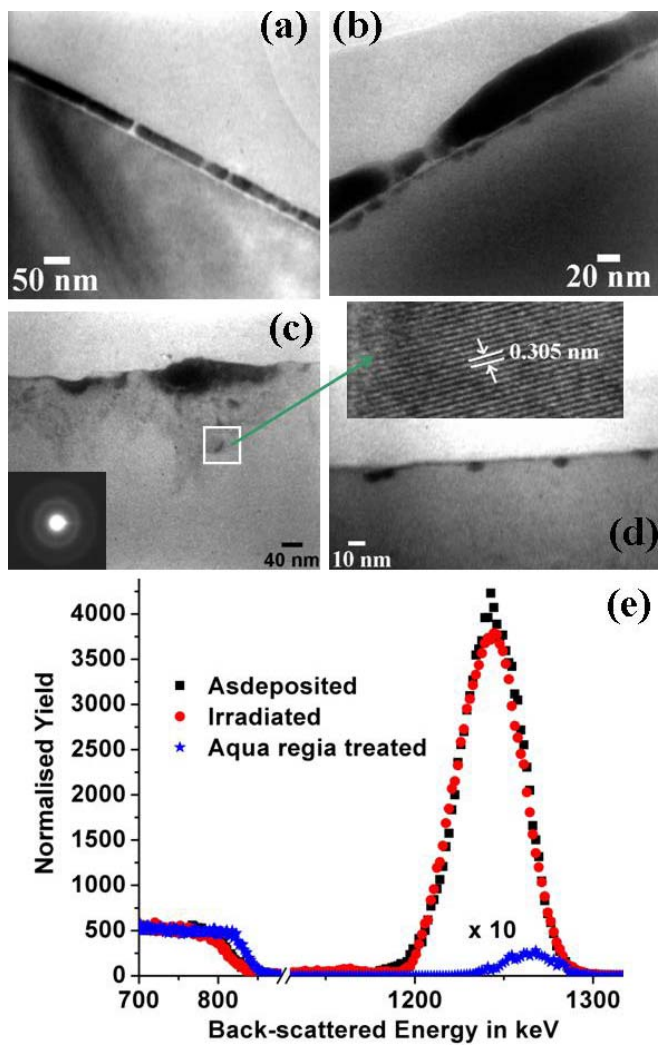

**Figure 3: J. Ghatak et al.**

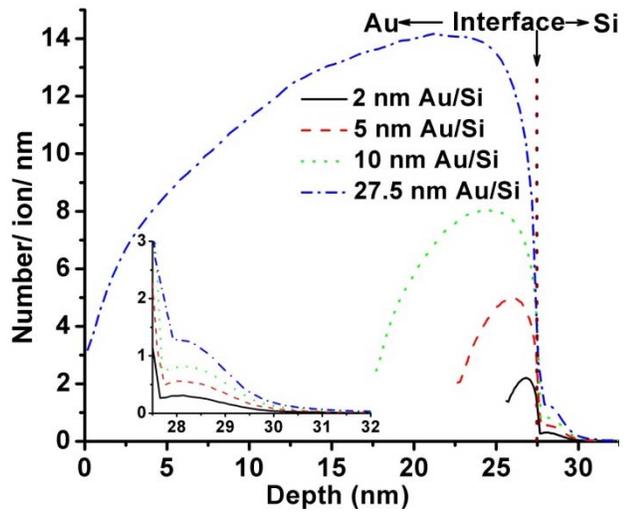